\documentclass[aps,prb,amsmath,amssymb,reprint,floatfix,showpacs]{revtex4-1}
\pdfoutput=1  
\usepackage{dcolumn}
\usepackage{graphicx,psfrag,epsf,epsfig}
\usepackage{dcolumn}
\usepackage{bm,bbm}
\usepackage{pstricks,pst-plot}     
\usepackage{latexsym}
\usepackage{mathrsfs,amsmath,amssymb,textcomp}
\usepackage{tikz}  
\usetikzlibrary{matrix,arrows} 
\usepackage{hyperref}
\usepackage{float}

\begin{document}


\title[]{Theoretical investigation of direct and phonon-assisted tunneling currents in InAlGaAs-InGaAs bulk and quantum
well interband tunnel junctions for multi-junction solar cells}

\author{U. Aeberhard}
\affiliation{ 
IEK5-Photovoltaik, Forschungszentrum J\"ulich, 52425 J\"ulich, Germany
}%


\date{\today}

\begin{abstract} 
Direct and phonon-assisted tunneling currents in InAlGaAs-InGaAs bulk and double quantum well interband tunnel
heterojunctions are simulated rigorously using the non-equilibrium  Green's function formalism for coherent and
dissipative quantum transport in combination with a simple two-band tight-binding model for the electronic
structure. A realistic band profile and associated built-in electrostatic field is obtained via self-consistent coupling
of the transport formalism to Poisson's equation. The model reproduces experimentally observed features in the
current-voltage characteristics of the device, such as the structure appearing in the negative differential resistance
regime due to quantization of emitter states. Local maps of density of states and current spectrum reveal the impact 
of quasi-bound states, electric fields and electron-phonon scattering on the interband tunneling current. In this way, 
resonances appearing in the current through the double quantum well structure in the negative differential resistance 
regime can be related to the alignment of subbands in the coupled quantum wells.
\end{abstract}

\maketitle

\noindent Interband tunnel junctions (TJ) are essential components in multi-junction solar cell devices, allowing for
the series connection of single junction subcells with different band gaps as used for an enhanced utilization of the  solar
spectrum \cite{yamaguchi:05,king:07}. In a recent publication, the use of a double quantum well (DQW) structure
in the junction region was shown to enhance considerably the peak tunneling current in an InAlGaAs-InGaAs TJ, at moderate loss
of transparency \cite{lumb:12}. For a proper analysis of the potential of this approach and to enable further
optimization of the junction design, an accurate and comprehensive theoretical description of the physical processes
involved in the transport is required. However, conventional models for tunnel junction as used for the simulation of
multi-junction solar cells are usually limited in this respect by  a large number of simplifying assumptions on the
density of  states participating in the tunneling, the occupation of these states and the assisting scattering
mechanisms \cite{hermle:08,hauser:10}.

In this paper, the tunnel junction is simulated using a rigorous quantum
transport approach based on the non-equilibrium Green's function (NEGF) formalism as used in the nano-electronics
community for the simulation of interband tunneling transistors \cite{ogawa:00_2,rivas:01,rivas:03,luisier:10,knoch:10}.
In combination with atomistic electronic structure theory and self-consistent coupling to a Poisson solver, the approach is able to provide
a realistic picture of the relevant density of states and associated current flow under non-equilibrium conditions, by 
accurately reflecting the effects of internal fields, carrier confinement and elastic as well as inelastic scattering mechanisms.

\begin{table}[b]
\caption{\label{tab:parameters} Material parameters used in simulations}
\begin{ruledtabular}
\begin{tabular}{lccl}
&In$_{0.52}$Al$_{0.33}$Ga$_{0.15}$As &In$_{0.53}$Ga$_{0.47}$As \\
\hline
 $E_{s}$ [eV]& 1.04 & 0.74 \\
 $E_{p}$ [eV]&  -0.14& 0.0  \\ 
 $V_{sp}$ [eV]&2.54& 2.6\\
$m^{*}_{el}/m_{0}$& 0.065 & 0.041\\
$m^{*}_{lh}/m_{0}$& 0.087 & 0.052\\ 
$\epsilon_{r}$&13.2&14.2\\
$\epsilon_{\infty}$&10.7&11.0\\  
$\hbar\Omega_{LO}$ [eV] &0.038&0.033\\ 
$D_{ac}$  [eV]&5&5
\end{tabular}
\end{ruledtabular} 
\end{table}  

The present approach is based on the implementation of the NEGF formalism for III-V quantum well (QW) structures with
photvoltaic applications as introduced in Ref.~[\onlinecite{ae:prb_08}], with the material parameters of the
nearest-neighbor two-band tight-binding (TB) model and interactions adjusted to the InAlGaAs-InGaAs system of
Ref.~[\onlinecite{lumb:12}], as presented in Tab.~\ref{tab:parameters}. The orbital basis, in difference to a band basis
like $\mathbf{k}\cdot \mathbf{p}$, enables a consistent and unified description of electrons independent of their 
position in energy with respect to conduction and valence bands and is thus ideally suited for interband tunneling
processes. The TB parameters are adjusted to reproduce the zone center band gaps and effective masses of the bulk 
component materials, choosing the light hole mass for the description of the valence band. Among the various existing 
scattering mechanisms for electrons and holes, elastic coupling to acoustic phonons is considered within the deformation
potential approximation, and inelastic interaction with polar optical phonons is included via the corresponding
Fr\"ohlich 	Hamiltonian, with parameters also displayed in Tab.~\ref{tab:parameters}.
While the former process results in a broadening of the spectral quantities in both energy and
transverse momentum domain, the latter 	is essential for the description of the phonon mediated
tunneling current at vanishing overlap of electron and hole states in the injector regions.  At the
present stage, no trap-assisted tunneling relying on additional defect states is included, which
potentially leads to an underestimation of the dark interband current, especially at higher forward bias voltages.

\begin{figure}[t]
\begin{center}
\includegraphics[width=8cm]{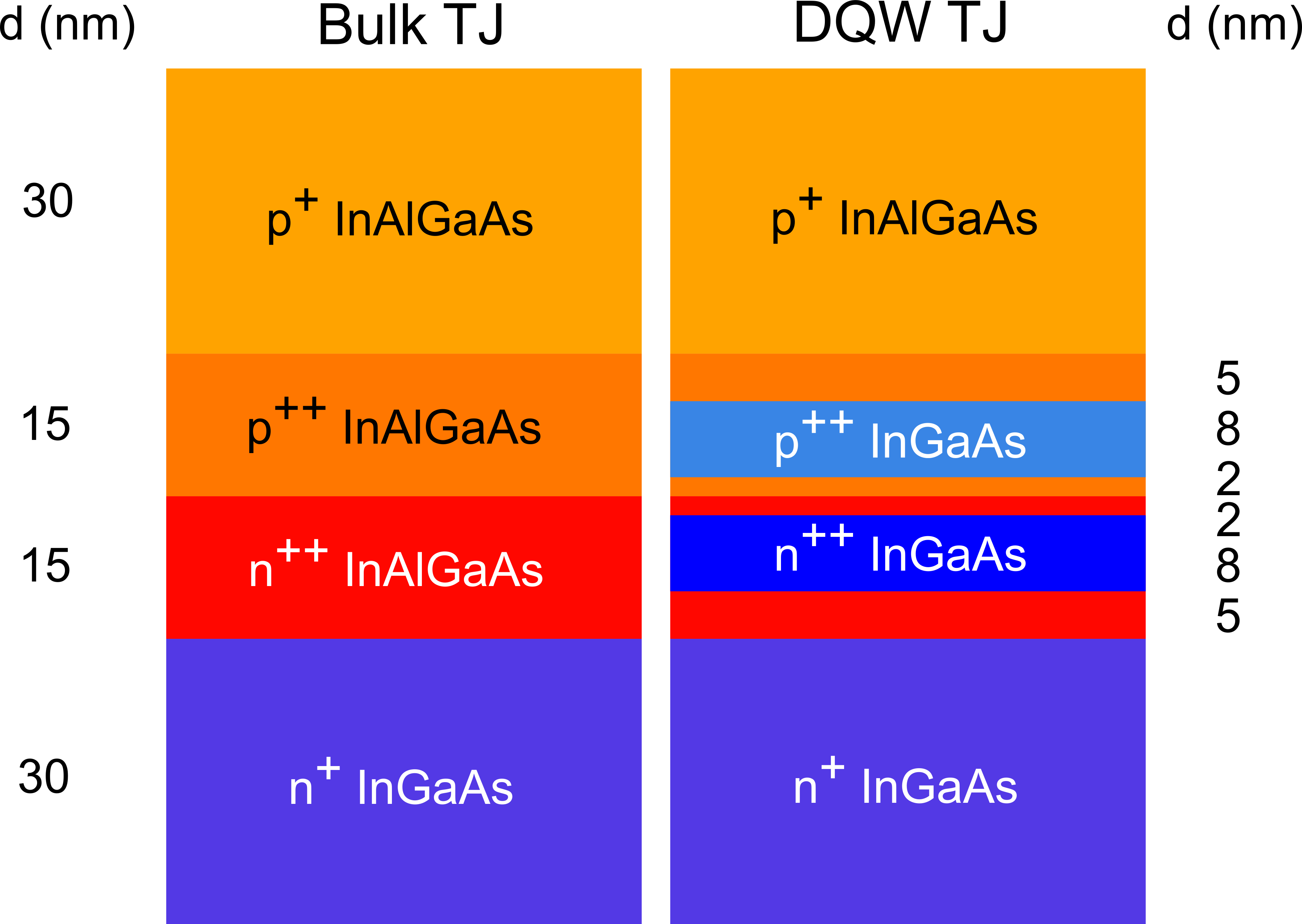} 
\caption{Schematic layer and material structure of the bulk and DQW tunnel-heterojunction devices from Ref.
[\onlinecite{lumb:12}]. The doping density amounts to $N_{d/a}=2\times 10^{17}$ cm$^{-3}$ for the
lightly doped $n^{+}/p^{+}$ regions and $N_{d/a}=10^{19}$ cm$^{-3}$ for the heavy $n^{++}/p^{++}$ doping.
\label{fig:structures}}
\end{center}
\end{figure}
\begin{figure}[t]
\begin{center}
\includegraphics[width=7cm]{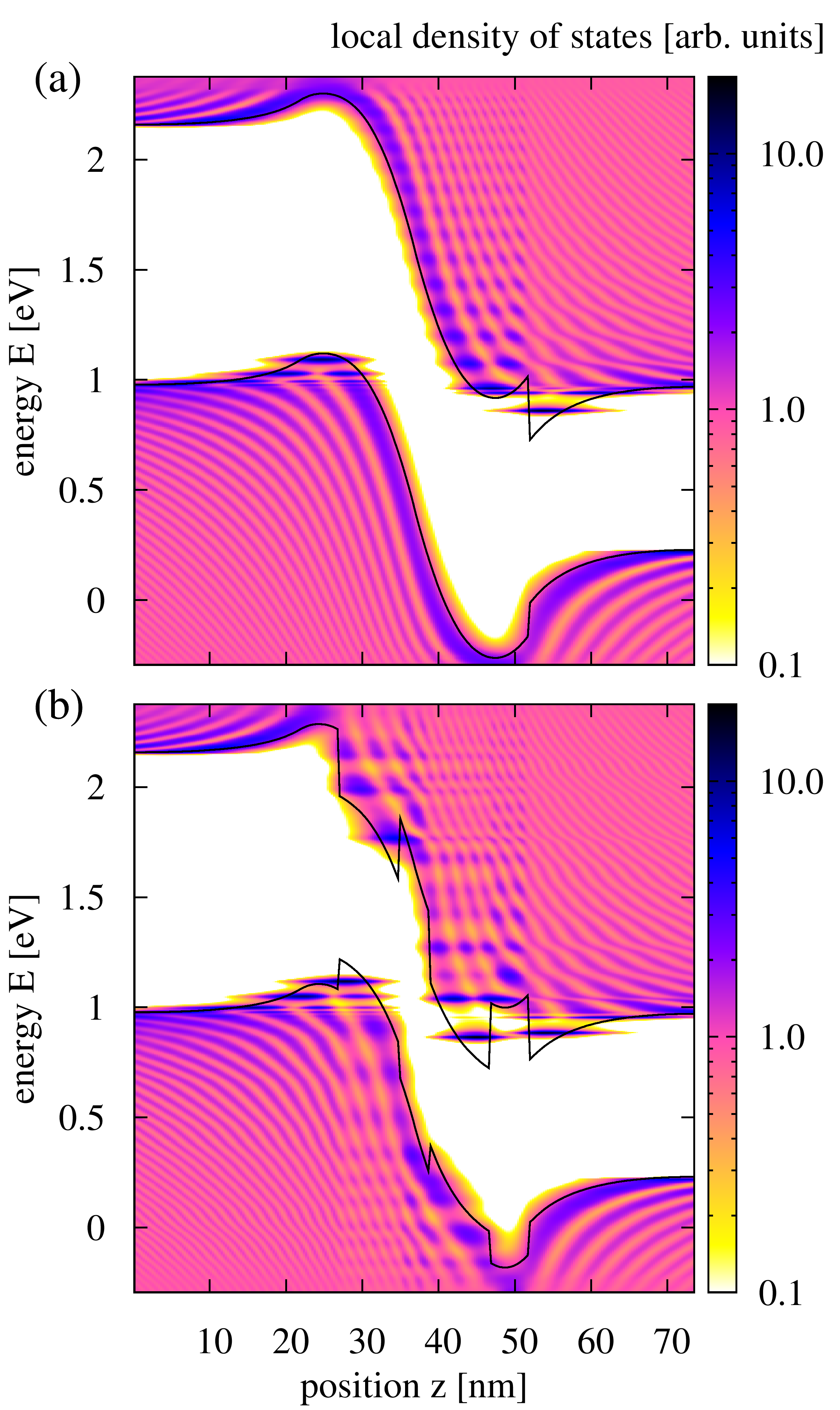}  
\caption{Band profile and local density of states at vanishing transverse momentum and zero
bias voltage, for (a) the bulk heterojunction device, exhibiting state quantization in the highly doped emitter
regions adjacent to the junction and (b) the DQW heterojunction device, where the strong band bending leads to the
unbinding of the high-lying confined states.
\label{fig:ldos}}
\end{center}
\end{figure}
\begin{figure}[t]
\begin{center}
\includegraphics[width=6cm]{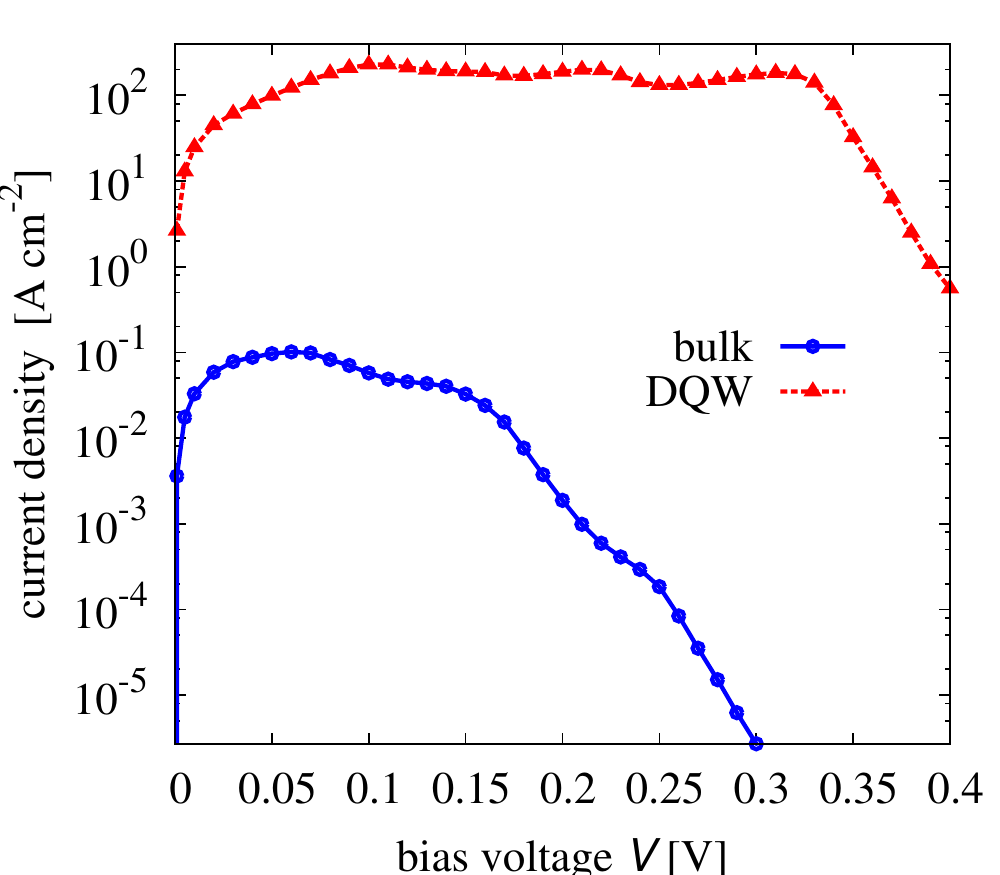}  
\caption{Current-voltage characteristics of bulk and DQW heterojunction devices at forward bias.
Due to the lower effective barrier potential, the peak current in the DQW device exceeds that of
the bulk junction by more than two orders of magnitude and occurs at much larger bias
voltage. The structure in the negative-differential resistance regime exhibited by the bulk device
can be ascribed to the quasi-2D state quantization in the emitter regions adjacent to the junction,
as shown in Fig.~\ref{fig:ldos}. The corresponding features in the IV-characteristics of the DQW
heterojunction 	device are considerably stronger and are identified as the signatures of resonant
tunneling between QW states.
\label{fig:iv}}
\end{center}
\end{figure}

The device structures under consideration are displayed in Fig.~\ref{fig:structures} and
correspond to the central region of the devices discussed in Ref.~[\onlinecite{lumb:12}]. The tunnel
junction consists of two adjacent 15 nm thick layers of
In$_{0.52}$Al$_{0.33}$Ga$_{0.15}$As with high doping level of $N_{d/a}=10^{19}$ cm$^{-3}$. In the
DQW structure, 8 nm of In$_{0.53}$Ga$_{0.47}$As are inserted on each side of the
junction to form the potential wells leading to a local constriction of the interband tunneling 
barrier. For the present simulation, the size of the lightly doped ($N_{d/a}=2\times 10^{17}$
cm$^{-3}$) buffer layers on top and bottom of the tunnel junction is reduced to 30 nm, and
neither the p-doped InGaAs capping layer nor the n-doped InP substrate present in the original
structures are included.
   
The equilibrium band edge profiles and associated local density of states (LDOS) at
vanishing transverse momentum  ($\mathbf{k}_{\parallel}=0$) corresponding to the above layer structures and doping
levels are shown in Fig.~\ref{fig:ldos}. In the bulk heterojunction case, displayed in Fig.~\ref{fig:ldos}(a), the
strong band bending close to the junction results in the appearance of bound states in both electron and hole emitter regions. While these states
feed the tunneling current, they are themselves not accessible from the contacts without energy relaxation. Hence, there is only a vanishingly narrow forward bias
 regime with finite energetic overlap of bulk electrode states at the low doping levels of the n$^{+}$ and
 p$^{+}$ regions, where ballistic transport from one contact to the other via exclusively direct tunneling in the
 junction region is possible, in spite of the lower band gap at the $n$-contact.  As a consequence, any description
 based on a ballistic picture of direct tunneling will severely underestimate current flow even close to zero bias. 
 There is, however, an extended forward bias regime with finite energetic overlap of interface states close to the
 junction via which the tunneling can proceed.  In the DQW case, displayed in Fig.~\ref{fig:ldos}(b), both the energetic
 overlap and spatial extension of interface states is increased due to the presence of the additional confined subgap
 states in the quasi two-dimensional QW regions close 	to the junction. The extremely strong band bending results in a partial
 unbinding of the higher QW states, such that the LDOS is no longer well described by the states of a  square well potential, but rather a new confinement
 potential 	reflecting the effects of both band bending and band offsets. In combination with the confined states at the
 heterojunction interface, the injector regions thus provide a complex combination of bound, quasibound and continuum 
 states participating in the interband tunneling process.

\begin{figure}[t]
\begin{center}
\includegraphics[width=8.5cm]{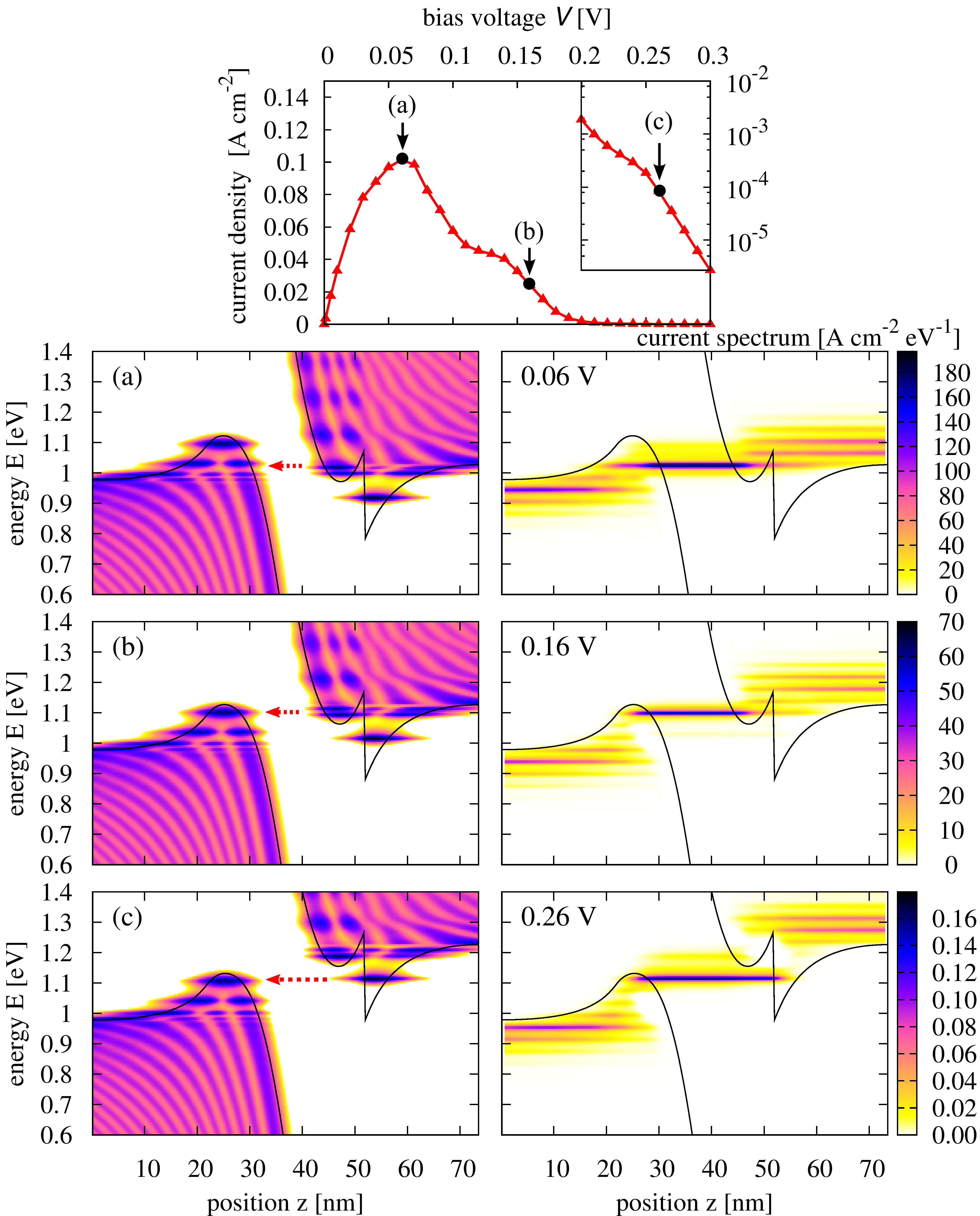}  
\caption{Current-voltage characteristics, local density of
states and corresponding local current spectrum of the bulk tunnel-heterojunction device at forward
bias voltages of (a) 0.06 V, (b) 0.16 V and (c) 0.26 V, illustrating different interband tunneling
processes depending on the energetic alignment of localized and extended states in the injection regions adjacent to the junction.
\label{fig:current_bulk}} 
\end{center}
\end{figure} 

 \begin{figure}[t]
\begin{center} 
\includegraphics[width=8.5cm]{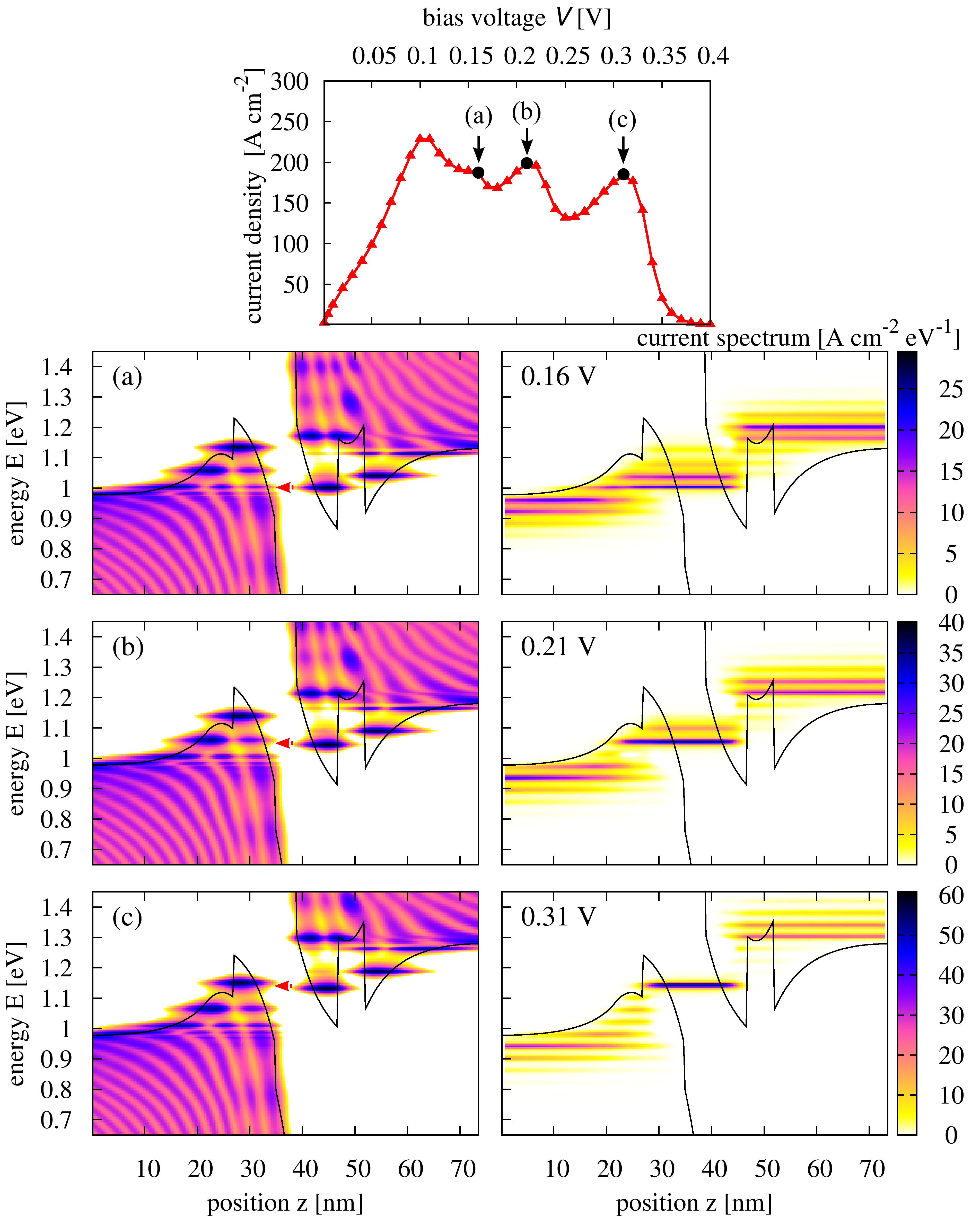}  
\caption{Current-voltage characteristics, local density of states and corresponding
local current spectrum of the DQW tunnel-heterojunction device at forward bias voltages of (a) 0.16
V, (b) 0.21 V and (c) 0.31 V, at which electrons tunnel resonantly from the lowest QW subband at the n-side
of the junction to the third, second and first QW subband at the p-side, respectively.
\label{fig:current_dqw}}
\end{center}
\end{figure}
 
 Fig.~\ref{fig:iv} displays the current-voltage characteristics of the interband charge
 transport in the bulk and DQW heterojunctions as computed with the NEGF model. The current-voltage
 characteristics reproduce the qualitative features of the experimental data in Ref.~[\onlinecite{lumb:12}] in terms of 
 significantly higher peak current and peak voltage, resulting from the lower effective barrier
 potential, but lower negative differential resistance (NDR) for the DQW TJ as compared to the bulk
 TJ. As can be inferred from Fig.~\ref{fig:ldos} (a), the structure appearing in the NDR region of 
 the IV characteristics for the bulk heterojunction is due to discretization of the longitudinal DOS
 in the injection regions adjacent to the junction. In the DQW sample,  the local current maxima are
 much more pronounced and are identified as the signatures of resonant tunneling between different QW states.
 
 In order to analyze the current flow in the two structures in more detail with respect to the different extended and
 localized states involved, the current spectrum at different bias voltage is computed for the full spatial extent of
 the device. In the case of the bulk heterojunction, the result displayed in
 Fig.~\ref{fig:current_bulk}, together with the corresponding LDOS ($\mathbf{k}_{\parallel}=0$),
 allows the distinction of different regimes of interband charge transport depending on the applied bias:
 (a) At the peak voltage of 0.06 V, current flow is dominated by carriers first  tunneling through the triangular
 barrier of the hetero-interface at the electron contact into the lowest subband of the doping-induced potential well
 at the n-side of the junction and then into the second subband of the doping-induced potential well at the
 p-side, from where escape to the hole contact proceeds via electron-phonon scattering. A minor contribution is due
 ballistic transport above the triangular barrier into the subband associated with the lowest localized state at the
 p-side; (b) At a bias of 0.16 V, after initial relaxation close the contacts mediated by multiple scattering events with phonons, current flow
 proceeds again predominantly via tunneling through the triangular barrier from the n-side, but now into the lowest
 subband of the doping-induced potential well at the p-side; (c) At even larger bias voltage of 0.26 V, the major part
 of the current flows via the lowest subband of the triangular potential well at the hetero-interface into the lowest
 subband at the p-side, overcoming a substantially thicker tunneling barrier, while only a small fraction of carriers
 are transported via phonon-assisted tunneling from the higher interface states.  Beyond this transport regime, i.e., for
 considerably larger separations of contact Fermi levels, there are no more available current paths in terms of
 energetically overlapping states in the injection regions, and current breaks down exponentially, until thermionic
 emission sets in, or, in more realistic cases, defect-assisted tunneling takes over. In
 the DQW TJ, as shown in Fig.
 \ref{fig:current_dqw} the dominant role in the transport process is played by the QW states, which in case of energetic
 alignement at specific values of the bias voltage give rise to resonant tunneling. On the n-side, current flow proceeds
 over the whole bias regime via tunneling into the second subband of the QW and subsequent relaxation into the lowest 
 subband, while the states in the triangular well at the heterointerface seem to be irrelevant for transport. On the
 p-side, there is a strong bias dependence of the specific subband involved in the interband tunneling process: (a) at
 0.16 V, the alignement is between the lowest subband at the n-side and the third subband state at the p-side, (b) at
 0.21 V, the second subband is resonant and (c) at 0.31 V, carriers are transferred between the lowest subbands of 
 the QW. Again, beyond this point, energetic overlap is limited to phonon sidebands, which results in an exponential
 decay of the current with increasing forward bias voltage. Finally, in both tunnel junction
 devices, due to the prominent role of narrow bands with enhanced transmission, the current spectrum
 close to the contacts shows a pronounced structure depending on the phonon energy, in spite of the
 bulk-like, continuous DOS in these regions.
 
In conclusion, we have presented a rigorous approach to the simulation of carrier transport in
interband tunnel junctions with applications in multi-junction solar cells, which is able to account for both
direct and phonon-assisted contributions under consideration of realistic spatial potential
variations, including arbitrarily shaped quantum well structures. By consideration of the local maps
of DOS and current spectrum, a microscopic picture of the current flow at different values of
forward bias voltage is obtained. Apart from providing insight into the dominant transport
mechanisms at given bias, the simulation results are in qualitative agreement with the
experimentally found large increase in peak current and peak voltage as well as a decrease of NDR by
the insertion of a double quantum well structure in the junction region. The quantitative agreement
of the tunneling currents with the measured values is expected to be improved by the use of a larger
tight-binding basis such as $sp^3d^5s^{*}$ \cite{jancu:97}, a more accurate treatment of the
scattering processes and, most notably, the inclusion of defect-assisted tunneling processes. The
latter is essential to extend the present approach to multi-junction solar cell devices based on
disordered and defect-rich materials, such as the micromorph silicon thin-film tandem solar cell.
     
\bibliographystyle{apsrev4-1}
\bibliography{/home/aeberurs/Biblio/bib_files/tunnel_junction,/home/aeberurs/Biblio/bib_files/aeberurs,/home/aeberurs/Biblio/bib_files/negf,/home/aeberurs/Biblio/bib_files/pv,/home/aeberurs/Biblio/bib_files/bandstructure_TB}%
 
\end{document}